\documentclass[reprint,amsmath,amssymb,aps,pra,superscriptaddress]{revtex4-1}
\usepackage{amsfonts}
\usepackage{amssymb}
\usepackage{amsmath}
\usepackage{calc}
\usepackage{ulem}
\usepackage{subfigure}
\usepackage{graphicx}
\usepackage{epstopdf}
\usepackage{dcolumn}
\usepackage{bm}
\usepackage{color} 
\usepackage{txfonts}
\usepackage[dvipsnames]{xcolor}
\usepackage[colorlinks, citecolor=blue]{hyperref} 
\usepackage{ulem}
\usepackage{multirow}
\usepackage{amsmath}
\DeclareMathOperator*{\argmin}{argmin}

\newcommand{\beq}{\begin{equation}}  
\newcommand{\eeq}{\end{equation}}
\newcommand{\beqa}{\begin{eqnarray}}
\newcommand{\eeqa}{\end{eqnarray}}

\newcommand{\figref}[1]{\mbox{Fig.~\ref{#1}}}

\newcommand{\figpanel}[2]{Fig.~\hyperref[#1]{\ref*{#1}(#2)}}
\newcommand{\figpanels}[3]{Fig.~\hyperref[#1]{\ref*{#1}(#2)-(#3)}}
\newcommand{\figpanelNoPrefix}[2]{\hyperref[#1]{\ref*{#1}(#2)}}

\usepackage{mleftright} 
 
\begin{document}
\title{Optimal Control by Variational Quantum Algorithms}
\author{Tangyou Huang}
\email{htangyou@gmail.com}
\affiliation{Shanghai Qi Zhi Institute, AI Tower, Xuhui District, Shanghai 200232, Shanghai, China}

\author{Jing-Jun Zhu}
\affiliation{Laboratoire Interdisciplinaire Carnot de Bourgogne, CNRS UMR 6303, Universit\'e de Bourgogne Franche-Comt\'e, BP 47870, 21078 Dijon, France}

\author{Zhong-Yi Ni}
\affiliation{Hong Kong University of Science and Technology (Guangzhou), Guangzhou, China}
\date{\today }
	
\begin{abstract}
Hybrid quantum-classical algorithms hold great promise for solving quantum control problems on near-term quantum computers. 
In this work, we employ the hybrid framework
that integrates digital quantum simulation with classical optimization 
to achieve optimal engineering of quantum many-body systems. To evaluate the overall performance of this method, we introduce a general metric termed \textit{control optimality}, which accounts for constraints on both classical and quantum components.
As a concrete example, we investigate the time-optimal control for perfect state transfer in a one-dimensional spin model using the variational quantum algorithm, closely approaching the quantum speed limit. 
Moreover, we discuss the emergent gradient behavior and error robustness, demonstrating the feasibility of applying hybrid quantum algorithms to solve quantum optimal control problems. 
These results establish a systematic framework for hybrid algorithms to address quantum control problems on near-term quantum platforms.
\end{abstract}

\maketitle
\section{introduction}
The rapid development of quantum computing has opened unprecedented opportunities for addressing complex problems that are intractable for classical computers~\cite{feynman1985quantum,daley2022practical}. Although in the early stages of fault-tolerant quantum device development~\cite{PRXQuantum.5.020101}, significant progress has been made in applications such as material design~\cite{bauer2020quantum}, quantum cryptography~\cite{quantum_cryptography}, and solving optimization problems~\cite{farhi2014quantum}. 
Among these, quantum optimal control of many-body systems~\cite{Doria2011prl} appears to be a particularly challenging task, in which the exponential increase of Hilbert space and intricate interactions of a quantum many-body system inevitably pose the obstacles for their classical counterparts. 
Along this line, variational quantum algorithms (VQAs)~\cite{cerezo2021variational} 
emerge as a promising strategy to achieve practical quantum advantage by using quantum processors for complex simulation and classical optimization techniques to iteratively refine the effective variables. 
{While quantum optimal control techniques have been used to enhance the performance of VQAs in terms of pulse-level engineering and circuit ansatz design~\cite{magann2021pulses,deKeijzer2023pulsebased,QOCvqa}, the hybrid method can also be applied to solve quantum control problems such as optimal state preparation in nuclear magnetic resonance~\cite{Sun2017prl}, circuit learning for the time-optimal control of a trapped quantum particle~\cite{huang2023time}, nonadiabatic processes in molecular systems~\cite{Moleculardynamic,MolecularDQS}, and digital counter-diabatic driving~\cite{DAQC1}. In this study, we focus on solving the optimal control problem of a many-body system within the VQA framework.}

Generally, to implement this hybrid quantum-classical workflow, one shall first simulate the driving dynamic of a quantum system on the quantum processor, as opposed to conventional methods, i.e., such as the time-dependent density matrix renormalization group~\cite{Doria2011prl} and the tensor network approach~\cite{orus2019tensor}. 
Remarkably, digital quantum simulation relying on the Trotter-Suzkui formula~\cite{Trotter1} has been widely investigated theoretically and experimentally, indicating a quantum simulator can be used to mimic the dynamical behaviors of a complex quantum system, despite the gate imperfection and decoherence shown in the noisy intermediate-scale quantum (NISQ) devices ~\cite{preskill2018quantum,kim2023scalable,smith2019simulating,fauseweh2024quantum}. 
{In practice, there are several well-developed quantum simulators, i.e. analogy simulators, digital-analogy simulators, and digital simulators. In our case, we concentrate on the last one to solve the corresponding optimal control problem. To mimic the unitary evolution of an arbitrary Hamiltonian, typically, the quantum process comprises a large number of elementary quantum gates. The variables (controllers) of the Hamiltonian are then optimized by minimizing a given objective function using numerical algorithms such as gradient-involved (-free) optimization and machine learning protocols~\cite{torlai2020machine}.} 
On a NISQ device, simulation errors can occur due to the Trotter approximation being limited to a finite number of gates~\cite{smith2019simulating}. 
There are also decoherence errors, such as gate imperfections and qubit relaxation, as well as optimization problems such as the gradient-vanishing problem (the so-called Barren Plateau  phenomenon)~\cite{mcclean2018barren,wang2021noiseBP,cerezo2021costBP}, which hinder the practical implementation of VQA-based solvers for quantum optimal control. 
 
Regarding the optimization efficiency, previous works have analyzed the trainability of quantum algorithms by exploring the expressibility of circuit ansatze~\cite{sim2019expressibility} estimated by the divergence of fidelity distribution from the Haar measurement, and its connection to the Barren Plateau (BP) has been studied~\cite{nakaji2021expressibility}. 
In terms of the Trotter errors, recent studies have indicated that they might be overestimated in the digital quantum simulator ~\cite{yi2021pra,trotter2022prl,Self-Healing2023digital}, and they can be mitigated through an adaptive optimization~\cite{zijiang2023prl,Zhao2023prxq} or even by leveraging effective counter-diabaticity from Trotter decomposition in digital adiabatic quantum algorithms~\cite{Wurtz2022counterdiabaticity}. 
However, these studies lack a comprehensive framework for describing the overall performance of hybrid  workflows for combining both quantum and classical aspects.

In this {article}, we {fisrt provide} the theoretical background of VQAs as a hybrid method for solving quantum control problems, emphasizing its significance within the broader contexts of quantum computing and quantum optimal control. To evaluate the efficiency of this workflow, we introduce the concept of \textit{control optimality}, a metric that statistically quantifies the overall performance of hybrid quantum-classical algorithms across multiple dimensions, including circuit expressibility, optimizer efficiency, and Trotter error.
As a case study, we apply this framework to achieve time-optimal control for perfect state transfer in a one-dimensional spin chain within a shallow circuit. Furthermore, we analyze the gradient behavior and the robustness of the classical optimizer under state preparation and measurement (SPAM) errors.
Our results highlight the feasibility and effectiveness of the VQA-based approach in addressing quantum optimal control problems on NISQ devices.

\begin{figure}[h]
\centering
\includegraphics[scale=0.12]{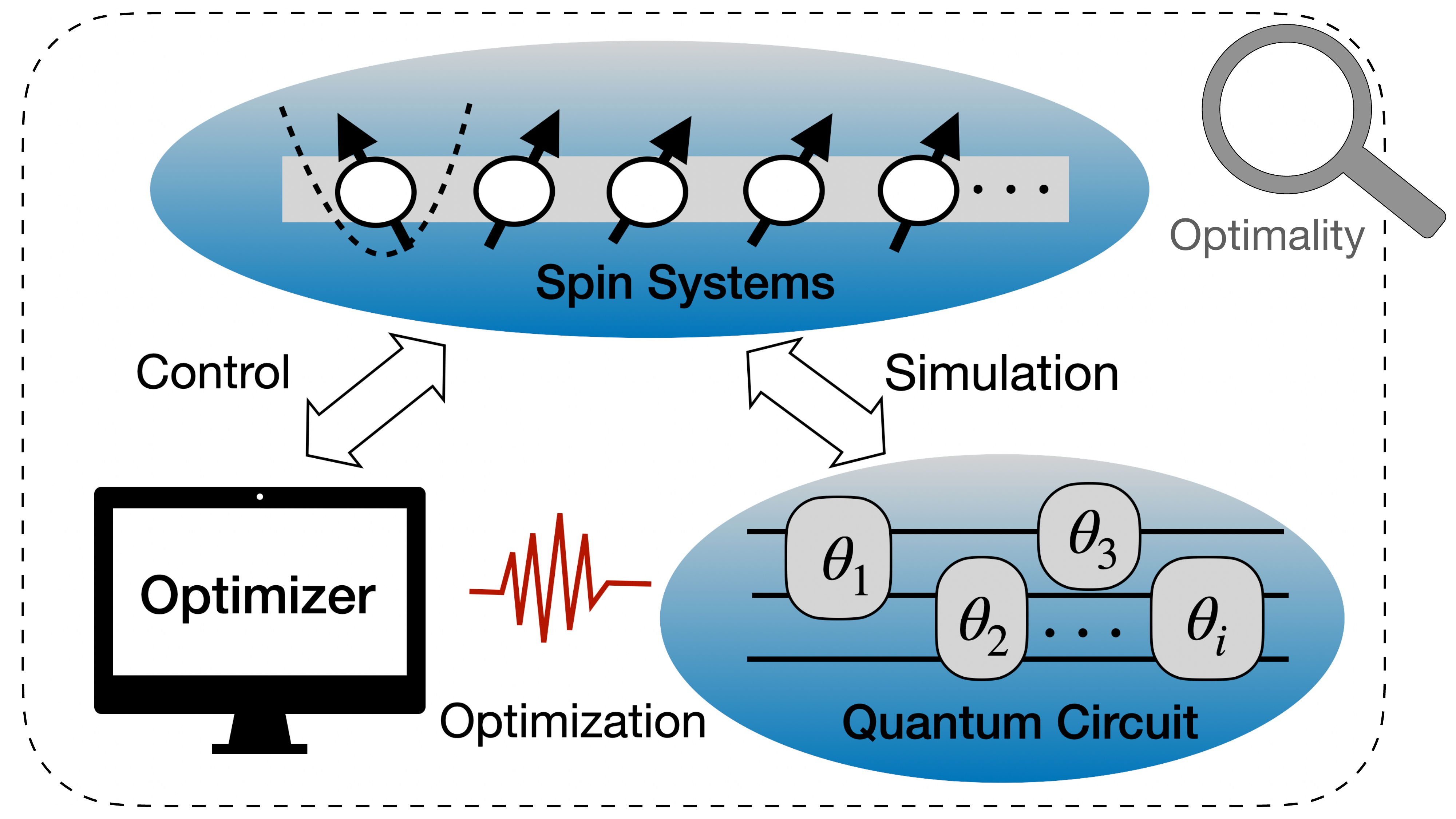}
\caption{The workflow diagram of quantum optimal control with a hybrid quantum-classical framework. }
\label{fig:figure1}
\end{figure}

\section{Background and notation}

\subsection{ Quantum optimal control}

The dynamics of a closed quantum system is governed by the time-dependent Schrödinger equation as follows (for simplicity, we set $\hbar\equiv 1$ through all the context): 
\beq\label{eq:TDSE}
i \frac{\partial|\psi\rangle}{\partial t}= \left[H_0+\sum_{\gamma=1}^{\Gamma}\lambda_\gamma(t)H_\gamma\right]|\psi(t)\rangle,
\eeq
where the wave function $|\psi(t)\rangle$, the static Hamiltonian $H_0$, and driving Hamiltonians $\{H_\gamma\}_{\gamma=1}^{\Gamma}$ are associated with controllers $\{\lambda_\gamma(t)\}^{\Gamma}_{\gamma=1}$, which are confined in a manifold $\lambda_\gamma(t)\in \mathbb{R}_\lambda$. In the following, we refer to the control assembled as a control policy $\Lambda(t) =\{\lambda_\gamma(t)\}^{\Gamma}_{\gamma=1}$.
For a given time $T$, the corresponding time-evolution operator is given by
\beq\label{eq:Ucont}
\mathcal{U}(0,T) = \mathcal{T}\exp\left(-i\int_{0}^{T}dt' \left[H_0+\sum_{\gamma=1}^{\Gamma}\lambda_\gamma(t')H_\gamma\right]\right),
\eeq
which transforms the initial state $|\psi(0)\rangle = |\psi_{\text{in}}\rangle$ to the final state $|\psi(T)\rangle = \mathcal{U}(0,T)|\psi(0)\rangle$ by an assigned control policy, where $\mathcal{T}$ denotes the time-ordering. We emphasize that the trajectory of the state is unique for a given control policy $\overline{\Lambda}$ with the initial condition $|\psi(0)\rangle = |\psi_{\text{in}}\rangle$. 
In order to complete the control task by preparing the target state $|\psi_{\text{tar}}\rangle$ from the initial state $|\psi_{\text{in}}\rangle$ , one needs to find the optimal control policy 
$\Lambda^{\text{opt}}$ that minimizes an objective function
\beq\label{OCcondition}
\Lambda_{\text{opt}}(T)= \argmin_{\Lambda\in\mathbb{R}_\Lambda }\mathcal{J}[T,\Lambda(t)],
\eeq
where the objective function reads
\beq\label{OTC_control}
\mathcal{J}(T,\Lambda) = \int_{0}^{\tau} \mathcal{J}[|\psi(s)\rangle ,\Lambda(s)]ds.
\eeq
Generally, the objective function is the fidelity between the target and final states,
\beq\label{fidelity}
F(\rho(T),\rho_\text{tar}) =  \text{Tr}\left(\sqrt{\sqrt{\rho_\text{tar}} ~{\rho}(T) \sqrt{\rho_\text{tar}}}\right)^2,
\eeq
where $\rho(t) =|\psi(t)\rangle\langle\psi(t)|$ represents the density matrix ($\rho_{\text{tar}}$  is the target density matrix), and $\text{Tr}[\cdot]$ denotes the trace operation. 
The above formulation can be addressed using the Pontryagin Maximum Principle (PMP)~\cite{OCT2021prxq} or numerical algorithms~\cite{crab2022review}. {Meanwhile,} it is crucial to analyze whether the target states (or unitaries) can be achieved under constrained control. 
The controllability of a quantum control problem depends on the control problem and the optimization strategy. 
The former aspect is governed by fundamental physics limits or instrumental challenges in experiments, described by the quantum speed limit and standard quantum limits, while the latter mainly relies on software, including control ingredients, optimization hardness, and information processing from measurements. In this work, we mostly {study} the theoretical limits from quantum speed limits and numerical constraints.

\subsection{The variational quantum algorithm}
A variational quantum algorithm (VQA)~\cite{cerezo2021variational} consists of a parametric quantum circuit (PQC) that includes a sequence of elementary quantum gates and a classical optimizer, which optimizes the gate parameters based on an objective function evaluated on qubits. 
Generally, a PQC can be decomposed into deterministic layers $V_l(\theta_0)$ and parametric layers $U_l(\theta_j)$:
\beq\label{HVA}
\textbf{U}(\Theta) =\prod_{l=1}^{L}\left[  V_l(\theta_0) U_l(\theta_l)\right],
\eeq
where the $j$-th deterministic layer is given by $V_l(\theta_0)$, and the trainable layer is represented by $U_l(\theta_l) = \prod_k e^{-i\theta^k_l P_k}$, with constant $\theta_0$ and free parameters $\theta_j$, respectively. The operators $P_k$ refer to elementary quantum gates, such as Pauli-typed rotating gates.
The performance of VQAs partially depends on the circuit ansatz, which defines the structure of the gate layout. Common ansatz in quantum algorithms include the Hardware-Efficient (HE) ansatz~\cite{Leone2024practicalusefulness} and the Hamiltonian Variational (HV) ansatz ~\cite{Park2024hamiltonian}. 
The former is typically used in quantum machine learning as a quantum neural network and is considered a 2-design~\cite{mcclean2018barren}, where variables across trainable layers are uncorrelated. The latter structure aims to mimic the dynamics of a given Hamiltonian for applications in quantum many-body problems~\cite{PhysRevA.92.042303}, where gate parameters are directly tied to an explicit Hamiltonian.
The main difference between these two types of ansatz lies in the correlation among trainable layers. The HE ansatz typically exhibits much less correlation compared to the HV ansatz. {Hence, the HV ansatz has been often utilized to improve the VQA application and prevents the gradient vanishing problems in VQA applications.}

Then, we consider a problem-inspired objective function associated with an observable $\hat{O}$:
\beq\label{objectivefun}
J_{ \hat{O}}(\theta) = 
\text{Tr} [\hat{O}\textbf{U}(\theta)\rho(T)\textbf{U}^\dagger(\theta)],
\eeq
where $\rho(T)$ represents the $N$-qubit state measured at $t=T$ and $\textbf{U}(\theta)$ defined in Eq.\eqref{HVA}. By combining a classical optimizer, the trainable layers can be optimized by minimizing the cost function~\eqref{objectivefun}, forming a standard VQA process:
\beq \label{OTC_angle}
\theta^{\text{opt}}= \argmin_{ {\theta} \in \mathbb{R}_{\theta} }J_{ \hat{O}}(\theta).
\eeq
Such a hybrid quantum-classical workflow can be used for solving quantum control problems \cite{huang2023time}. For instance, as illustrated in Fig. \figref{fig:figure1}, the quantum control problem of a spin system can be simulated by a pre-defined PQC, where the control policy is mapped onto a series of trainable gates optimized by the classical optimizer. Consequently, the optimal gate angles $\theta^{\text{opt}}$ obtained from the VQA can be reversely mapped to the optimal control policy $\Lambda_{\text{opt}}$ in the quantum control problem as described in Eq. \eqref{OTC_control}. {Next, we remark a few existing problems for the gradient-based VQA application and introduce a general metric term, called \it{control optimality}.}

\section{Quantum Optimal Control with VQAs}\label{VCL} 
To solve the quantum optimal control problem using the VQA-based approach, we follow these three steps:\\

\begin{enumerate}
\item \textbf{Initialization}:
We firstly initialize the control policy $\Lambda(0)=\Lambda_0$ and prepare the qubit-register with state $\rho(0)= \rho_{\text{in}}$.
\item \textbf{Circuit Construction:}
Secondly, we construct the PQC $\textbf{U}(\theta)$ formed by DQS of driving propagator (Eq.\eqref{eq:Ucont} )
\beq\label{eq:Utro}
\textbf{U}(\theta)
\equiv \prod_{j=1}^{N_t} \prod_{\gamma=1}^{\Gamma}\exp\left(-i H_\gamma\theta(\lambda_\gamma,t_j)\Delta t\right)V_j,
\eeq
 $V_j$ represents the evolution under the static Hamiltonian term and $\Delta t = T/(N_t-1)$ is the discrete time interval. The circuit consists of alternating layers of gates, representing the time evolution under various control Hamiltonians $ H_\gamma $
 with control-related angles $\theta(\lambda_\gamma,t_j)$.
\item \textbf{Optimization:} Thirdly, a classical optimizer updates the parameters $\theta\in\mathbb{R}$ by minimizing the infidelity between the final state $\rho (T)$ and the target state $\rho_{\text{tar}}$. The optimization target is:
\beqa
\theta^{\text{opt}}& =& \argmin_{\theta\in \mathbb{R}} \left[1-F(\rho(T),\rho_{\text{tar}})\right],
\eeqa
where $F(\cdot)$ is defined in Eq.(\ref{fidelity}) and the final state reads $\rho (T)=   \textbf{U}(\theta)\rho_0  \textbf{U}(\theta)^\dagger$.
\end{enumerate}

{In this paper, the Trotter errors for a total time $T$ evolution is quantified by norm-2 distance from the time-continuous propagator defined in Eq.\eqref{eq:Ucont}~\cite{trotter21prx}:
\beq\label{eq:trotter_error}
\xi=\| \mathbf{U}-\mathcal{U}(0,T)\|_2,
\eeq
where $\mathbf{U}$ refers the circuit ansatz of the first-order product formula~Eq.\eqref{eq:Utro} for the sake of simpicity. To pursue higher precision of DQS, one can take into account high-order decomposition, see Ref.~\cite{trotter21prx}.}

To minimize the infidelity, gradient-based and gradient-free methods are commonly employed in a hybrid quantum algorithm.
The gradient information over a trainable parameter $\theta$ with a Pauli-type gate can be either calculated relying on the parameter-shift rule~\cite{shift2018pra,analyticgrad2019pra} or numerical differentiation, both of which are widely used in the VQA community~\cite{cerezo2021variational}.
It's essential to mention that the Barren Plateau (BP) phenomenon, which hinders the training process, can be a significant challenge for a hybrid workflow. Specifically, the gradient of the trainable parameters with respect to the objective function may exponentially decrease as a function of qubit number. This phenomenon has been widely investigated in various aspects~\cite{mcclean2018barren,wang2021noiseBP,cerezo2021costBP}.
In \cite{mcclean2018barren}, \textit{McClean et al} present an estimation
\beq\label{BP_var}
 {\rm Var}[\partial_{\theta}J] \sim \mathcal{O}(2^{-2N}),
 \eeq
for a unitary 2-design ansatz where the circuit parameters are sufficiently random while the number of layers grows linearly as the qubit number $N$. 
In our setup, to study the variance of the control gradient, we randomly initialize the controller $ \Lambda(0)$ and numerically calculate the gradient with respect to  \cite{shift2018pra,analyticgrad2019pra}
\beq\label{grad_est}
\frac{\partial \mathcal{J}}{\partial \lambda_{\gamma,j}}= \frac{ \mathcal{J}(\lambda_{\gamma,j}+\Delta \lambda)-\mathcal{J}(\lambda_{\gamma,j}-\Delta\lambda)}{2 \Delta \lambda},
\eeq
where the small deviation $\Delta\lambda\ll\lambda_{\gamma,j}$ with the variable $\lambda_{\gamma,j}=\lambda_{\gamma}(t_j )$ refers $\gamma$-th controller at time $t_j$. To calculate the variance, we set both $\gamma,j$ are randomly sampled from $t_j\in\{0,\Delta t,2\Delta t...,(N_t-1)\Delta t\}$ and $\lambda_\gamma\in\{\lambda_0(t),\lambda_1(t),...,\lambda_{\Gamma}(t)\}$ with a uniform probability distribution.
In our setup, the gate angle of $j$-th trainable layer $U(\theta_j)$ is correlated by controller $\lambda(t_j)$ and qubit index,  as expressed in:
\beq\label{f_fun}
\theta_{\gamma,j}  = f[\lambda_\gamma( t_j),\textbf{n}],
\eeq
where the virtual spatial $\textbf{n}\in\{1,2,3,...,N\}$ denotes the index of qubits.
With the above mapping function (i.e., $\lambda\to\theta$), the quantum optimal control is transferred to a VQA application~\cite{shift2018pra,huang2023time}, where it leverages the quantum advantage~\cite{feynman2018simulating} since the dynamic of a many-body system is provided by a quantum simulator instead of a classical counterpart~\cite{Doria2011prl,orus2019tensor}.  

\subsection{The circuit expressibility}
The expressibility of a PQC $C(\theta)$ quantifies its capability of unitary generation, which can be indicated by the Hilbert Schmidt distance \cite{sim2019expressibility,nakaji2021expressibility} 
\beq
\mathbb{E}^{(t)}(C)=\bigg\Vert\int_{\text{Haar}}^{}(|\psi\rangle\langle\psi|)^{\otimes t}d\psi
-\int_{\Theta}^{}(|\psi_\theta\rangle\langle\psi_\theta|)^{\otimes t}d\theta\bigg\Vert_{HS},\nonumber\\
\eeq
where $\int_{\text{Haar}}^{}$ refers the integration over a state $|\psi\rangle$ in unitary $t$-design~\cite{hunter2019unitary}. In this study, we only consider the $2$-design.
And the parametric state is formed by circuit $C(\theta)$, e.g., $|\psi_\theta\rangle =U_C(\theta)|\psi_0\rangle$ with trainable $\theta\in\Theta$. 
If the arbitrary unitary can be uniformly sampled from the Haar measure  $\mathbb{U}_{\text{Haar}}$, one can get the maximum expressibility with distance $\mathbb{E}^{(t)}(C)=0$.
To estimate the expressibility of circuit $C$, on shall calculate Kullback–Leibler (KL) divergence between two probability distributions of fidelities~\cite{sim2019expressibility,nakaji2021expressibility}:
\beq\label{KL}
\textbf{Expr}(C)= D_{KL}\left[\mathbb{P}(F_C),\mathbb{P}_{\text{Haar}}(F)\right],
\eeq
where $\text{P}(F_C)$ and $\text{P}(F_{\text{Haar}})$ are sampled fidelity probability distributions.
More specifically, ${F}_C =  |\langle\psi_{\theta_2}|\psi_{\theta_1}\rangle|^2$ refers the fidelity of two final states with two randomly sampled free parameters $\{\theta_1,\theta_2\}\in\Theta$ of circuit $C$, e.g., $|\psi_{\theta}\rangle=U_C(\theta)|0\rangle$.
Meanwhile, ${F}_{\text{Haar}} =  |\langle\psi_{\text{Haar}} |\psi_{\text{Haar}}^{'} \rangle|^2 $ with $|\psi_{\text{Haar}}\rangle =  {U}_{\text{Haar}}|0\rangle$,
where ${U}_{\text{Haar}}$ is randomly drawn from 2-design manifold~\cite{mezzadri2006generate}.
And the probability density $\text{P}(F_{\text{Haar}})$ over fidelity space has an analytic expression
\beq\label{haar}
\text{P}_{\text{Haar}}(F)=(2^N-1)(1-F)^{2^N-2},
\eeq
in $2^N$-dimension Hilbert space~\cite{sim2019expressibility,nakaji2021expressibility}.
The ansatz $C$ with a smaller value $\textbf{Expr}(C)$ has higher expressibility since the $D_{KL}(p\|q) =0$ for two identical distribution $p$ and $q$.

VQAs aim to find the optimal solution, often requiring a highly expressive circuit ansatz to maintain the state controllability.
Consequently, particularly for a gradient-based optimizer, it requires a huge number of measurements to evaluate the exponentially decreased gradient information in the presence of the BP phenomenon. 
In other words, balancing the trade-off between trainability and expressibility in gradient-based VQAs remains a significant challenge to be explored~\cite{PRXQuantum.4.040335,Expressibility2022prxq,PRXQuantum.2.040337}.
{In parctice, the $D_{KL}$ refers access the feature of quantum componets, lacking the accessment of classical aspect. In order to evaluate the overall performance of such hybrid workflow, we next elaborate the concept of \textit{control optimality}.}
 
\subsection{ The control optimality}
The VQA, incorporating a non-convex objective function, produces a set of local extrema in the training landscape of the cost function $\{\mathcal{J}(\tilde{\Lambda}_i)\}$, where the corresponding controllers belong to a control set, e.g., $\tilde{\Lambda}_i\in\tilde{\mathcal{C}}$.
This controller set $\tilde{\mathcal{C}}$ includes all control policies that satisfy a given criteria on the cost value, for example, the fidelity criteria, $F(\tilde{\Lambda}_i)\ge F_c$ while the infidelity is used as the cost function.  
We then assume that such a control policy (fulfilling the above criteria) is sampled from the high-dimension distribution $\tilde{\mathcal{P}}(\Lambda)$, i.e., $\tilde{\Lambda}\sim\tilde{\mathcal{P}}(\Lambda)$ with the following probability distribution $\tilde{\mathbb{P}}_{\Lambda}$. One can define the reference probability distribution (PD) on the fidelity basis ${\mathbb{P}}_{\text{ref}}(F)$ as a step function: 
\beq\label{eq:refPD}
\text{Reference PD:}~{\mathbb{P}}_{\text{ref}}(F) \equiv [\![F\ge {F}_c ]\!],
\eeq
$[\![\cdot]\!]$ is the Iverson bracket, indicating the probability ${\mathbb{P}}_{\text{ref}}(F)=1$ while $F\ge{F}_c$. This reference PD indicates the conditional optimal control, where the resultant controller always fullfills the condition, e.g., $F\ge F_c$. Such a control set $\tilde{\mathcal{C}}$ refers to the set of \textit{conditional optimal solutions} for a quantum control problem. 

In practice, the control protocol hardly prepares the target with an unit fidelity, we take $F_{c}\to 1$ asymptotically.
In our framework, the VQA-based method produces diverse controllers due to different setups in a VQA, i.e., the randomly sampled initial controller, diverse circuit ansatz, or using different optimizers, forming the probability distribution (PD) of resultant fidelity $\mathbb{P}_{\text{lea}}(F)$.
To estimate the learning PD, we assign the learning iteration ($N_{\rm lea}$): 
\beq
\text{Learning PD:}~ \mathbb{P}_{\text{lea}}(F;N_{\rm lea}),
\eeq
{where $N_{\rm lea}$ represents the learning budget of optimization process. In a VQA application, the learning iteration is affected by both classical and quantum components, such as the circuit expressibility and optimizer efficiency. We therefore define the parameter $N_{\rm lea}$ as the cumulative count of loss function evaluations, which corresponds to the total number of final-state projective measurements in experiments. We here emphasize that $N_{\rm lea}=0$ implies the classic component has not been taken into account.}

\begin{figure}
\centering
\includegraphics[width=0.85\columnwidth ]{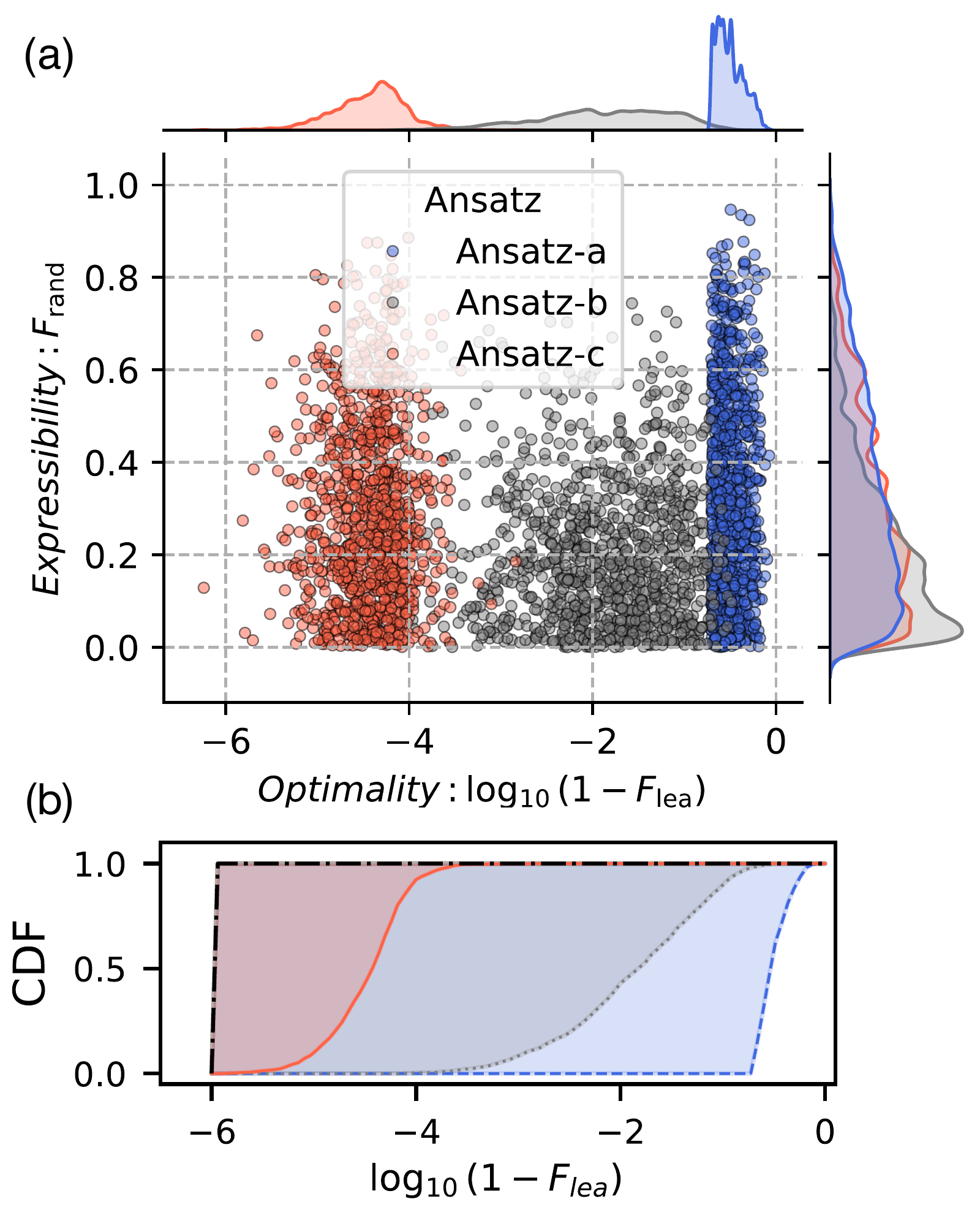}
\caption{(a) Each data point represents a single numerical experiment characterized by a pair of fidelity values, $(F_{\text{rand}}, F_{\text{opt}})$ based on a given control task, where $F_{\text{rand}}$ (the vertical axis) denotes the fidelity of a randomly sampled initial controller, and $F_{\text{lea}}$ (the horizontal axis) denotes the optimized fidelity. The horizontal axis corresponds to the optimized fidelity, while the vertical axis indicates the fidelity of the initial controller. The parameters used are $T = 4$, $N_t = 6$, and $N = 6$.
(b) The cumulative distribution functions (CDFs) of the fidelity distributions in (a) are shown for the reference (black dots), Ansatz-a (solid orange line), Ansatz-b (dotted grey line), and Ansatz-c (dashed blue line). The area between each CDF and the reference corresponds to the $W_1$ distance. Other parameters used in this experiment are the number of learning iterations $N_{\rm lea} = 10^4$.
}\label{fig:figure2}
\end{figure}

{Instead of measuring the distance from the random unitary, we here propose measureing the distance between the practical learning outcomes and the reference solution (Eq.\eqref{eq:refPD}) from the perspective of 
\textit{Empirical Information Theory}~\cite{PhysRev.106.620,majda2011improving}, which focuses on the behavior of information measures in practical, finite-sample settings.
In this vein, the \textit{optimality} of a given control problem is defined by calculating the distance between the empirical $\mathbb{P}_{\rm lea}(x)$ and reference probability distribution (PD)  $\mathbb{P}_{\rm ref}(x)$ by the one-dimensional Wasserstein ($W_1$) distance (also known as Earth Mover’s Distance)~\cite{panaretos2019statistical}:
\beq\label{eq:W1}
\textbf{Optim}  = W_1[\mathbb{P}_{\text{lea}}(x),\mathbb{P}_{\rm ref}(x)],
\eeq
where $W_1[\cdot]$ is the Wasserstein distance~\cite{panaretos2019statistical}. This quantity statistically estimates the overall performance of VQA-based optimal control with respect to various effects from the perspective of quantum optimal control theory~\cite{optimality1998pra,rabitz2004quantum,OCT2021prxq,huang2025quantumprocesstomographydigital}. It's worth noting that the definition $\textbf{Optim}$ is generic for an arbitrary cost function.
In the one-dimensional limits, the $W_1$ distance can be expressed as~\cite{panaretos2019statistical,huang2025quantumprocesstomographydigital}:
\beq
W_1(\mathbb{P}_{\rm lea}(x),\mathbb{P}_{\rm ref}(x))
=\int_{\mathbb{R}}^{} |\mathcal{D}_{\rm lea}(x)- \mathcal{D}_{\rm ref}(x)|dx,
\eeq
where $\mathcal{D}_P(x')=\text{Prob}[x \le x']$ denotes the cumulative distribution function (CDF) of the probability distribution $P$, representing the probability that the random variable $x$ takes on a value less than or equal to $x'$. An example is shown in \figref{fig:figure2}(b).
}

{We take the logarithm of infidelity as the basis $x=\log_{10}(1-F)\in[-\epsilon,0]$ bounded by a real and positive constant $\epsilon$.
For convenience, we set the reference PD as a delta function $ \mathbb{P}_{\rm ref} = \Delta(-\epsilon)$ at $x=-\epsilon $ referring to the optimal policy that always prepares the target state with the fidelity ${F}_0 = 1-10^{-\epsilon}$. In this way, the normalized $W_1$ distance is bounded by
$W_1/\epsilon \in[0,1]$  (seen in the Appendix \ref{app2})
\beq
0\le \textbf{Optim}[C( \Lambda)]\le 1,
\eeq
where $\textbf{Optim}[C( \Lambda)]=0$ represents the system is fully controllable while the distance $W_1[\mathbb{P}_{\text{lea}}, {\mathbb{P}}_{\rm ref}]=0$, indicating the control trajectory can be always optimized to an conditional optimal solution $\Lambda_{\text{opt}}\in\tilde{\mathcal{C}}$.
Larger distance $W_1[\mathbb{P}_{\text{lea}},\mathbb{P}_{\text{ref}}]\gg 0$ denotes the underlying control setup is less optimality. 
}

The optimality of the VQA reflects the overall performance of a  hybrid quantum-classical workflow, encompassing quantum aspects such as circuit design and error interference as well as classical aspects like the optimization landscape and robustness against the noise. 
To evaluate the control optimality of a hybrid algorithm, one must calculate a set of fidelity outcomes $\{F^{(i)}\}$ obtained based on a given learning budget $N_{\text{lea}}$. Note that optimality evaluations do not necessarily correspond to extrema of the cost-function landscape; in other words, convergence of the cost function is not a requirement for calculating the $\mathbb{P}_{
\text{lea}}$ (see discussion in Appendix B). We next evaluate control optimality under various control strategies in VQA-based quantum optimal control and make comparison with the results obtained in terms of the expressibility and Trotter errors.

\section{Models and Control setup}

To begin with, we consider one-dimensional Heisenberg model of $1/2$-spin chain , where the Hamiltonian reads
\beq\label{Heisenberg}                                                                        
{H}_{\text{XXZ}}= -\frac{1}{2}\sum_{n=0}^{N-2}J_0(\sigma^x_n\sigma^x_{n+1} +\sigma^y_n\sigma^y_{n+1})+ \sum_{n=0}^{N-1}B(t)\sigma_n^z.                          
\eeq
where $\sigma_n=\{ {\sigma}^x_{n},\sigma^y_{n},\sigma^z_{n}\}$ are Pauli operators acting on the $n$-th site. Hereby, $J_{0}$, $B_{n}(t)$, $N$ denote exchange interaction between nearest-neighboring sites, time-dependent driving field and the site number, respectively. {For simplicity, all the physical parameter are scaled in terms of $J_0$ $(J_0\equiv 1)$ to be dimensionless throughout the context.} Since the conservation of z-component magnetization,  i.e., the commutator $[ H_{\text{XXZ}},\sum_{n=1}^{N}\sigma^z_n]=0$, we restrict our analysis to single-excitation transport only. 

The single-excitation state can be described by $|\psi\rangle = \sum_na_n|\textbf{n}\rangle$, where $|\textbf{n}\rangle$ refers the state of the spin at $n$-th site is up (all the other spins are down).
For a given time $T$, our goal is to prepare the target state $|\psi_{\text{tar}}\rangle =|\textbf{N}\rangle $ from an initial state $|\psi_{\text{in}}\rangle=|\textbf{1}\rangle$, i.e., to transport a spin-up state from the first site to the final site of the chain. The perfect state transfer (PST)  along an one-dimension spin chain has been theoretically and experimentally investigated for quantum communication and fundamental physics like quantum speed limits (QSL)~\cite{MH06pra,chapman2016experimental}.

To reach this speed limit, the external field $B(t)$ can be designed as various kinds,
such as a parabolic potential~\cite{gong08pra}, P$\ddot{o}$chl-Teller potential~\cite{Tellerpotential} or a square well~\cite{squarewall}.
In this study, we focus on the harmonic trap 
\beqa\label{parabolic}
B_1[d(t),A(t)]=-\frac{1}{2}A(t)\sum_{n=1}^{N} \left[ {x}_n-d(t)  \right]^2 , 
\eeqa
where controller $A(t)$ is trap strength, $d(t)$ is the center of trap, and $x_n\in\textbf{x}$ with the virtual spatial $\textbf{x}=\{1,2,...,N\}$. 
Under the above parabolic design, several control problems have been studied in the past few years. These two controllers can be optimized simultaneously~\cite{ocqsl09prl,krotov2011pra}, or one can optimize one of them and set other invariable~\cite{ocqsl09prl,krotov2011pra,RLzhang2018}.


On the other hand, instead of a well-designed potential, we set the correlation-free driving field ,
\beq\label{free}
B_2(t) = \{\theta_n(t)\}_{n=1}^{N},
\eeq
where $\theta_n(t)$ is a variable of z-rotating angle $ {\tt R_z(\theta_n)} = e^{-i\theta_n\sigma_n^z}$ on $n$-th qubit. In other words, the state is evolved by $N$ independent time-varying control functions; for example, the driving term becomes $\sum_nB_2(t,n)\sigma_n^z$.

{For a given control function, the corresponding Hamiltonian Variational (HV) ansatz reads
\beq\label{eq: HVA}
\textbf{U}_{\rm HVA} = \prod_{l=1}^{N_t} \left(  U_{0} \cdot \bigotimes_{n=1}^{N} R_{z}(\theta^{(1/2)}_{n,l}) \right),
\eeq
where the unparametric layer $U_{0}= e^{-iJ_0(\sigma^x_n\sigma^x_{n+1} +\sigma^y_n\sigma^y_{n+1})T/N_t}$ refers the static term in Hamitonian \eqref{Heisenberg}, and variational parameter $\theta^{(1/2)}_{n,l}\sim B_{1/2}(t)$, respectively.
Next, we focus on the PST problem using the hybrid workflow for three control strategies (represented by three different ansatz):
\beqa\label{eq: three ansatz}
\textit{Ansatz-a}&:& B_1[d(t),A_0];\\
\textit{Ansatz-b}&:& B_1[d(t),A(t)];\\
\textit{Ansatz-c}&:& B_2(t);
\eeqa
resulting in various parametric layers $\textbf{U}_{\rm HVA}$ defined in Eq.~\eqref{eq: HVA}. One can find details on circuit construtions in the Appendix \ref{app1}. In a hybrid framework, when the dynamics under these three strategies are simulated using a digital quantum simulator, the aforementioned control strategies manifest as correlations among parameteric gates within the quantum circuit. 
In what follows, we combine one specific optimal control task, i.e. the maxiumum-fidelity control to analyze the performance of three above-mentioned ansatz in terms of the circuit expressibility and the control optimality, respectively.
}

\subsection{The maximum-fidelity control}

In this section, we formulate the VQA-based protocol to achieve maximum-fidelity control for a given time $T$. As an example, we demonstrate maximum-fidelity control for the PST problem with a spin number of $N=6$, total time $T = 4$, Trotter steps $N_t=12$, and a cost function $\mathcal{J} = 1 - F$, where the fidelity $F$ is defined in Eq.~(\ref{fidelity}).
To avoid physical limitations, the total time is chosen to be significantly larger than the minimum time required (see Eq.\eqref{time_minimum}) in this numerical experiment. The procedure begins with a random guess for the initial controller, and the same classical optimizer ({\tt SLSQP}\cite{SLSQP}) is used throughout. In this framework, the overall performance of the VQA protocol primarily depends on the control landscape of the ansatz, which includes circuit expressibility and optimizer capability. This can be effectively characterized by the control optimality metric.

In the two-dimensional distribution plot shown in \figref{fig:figure2}, we compare the behavior of control optimality and circuit expressibility across three ansatz. 
More specifically, we first randomly sample $10^3$ pairs of initial controllers and compute their corresponding fidelities $F_{\rm rand}$ (plotted on the y-axis of \figref{fig:figure2}).
Secondly, we get the resultant controller by learning the above initial setup (with $N_{lea}=20$ iterations) and get the distribution of logarithmic infidelity $\log_{10}(1-F_{\rm lea})$ (plotted on the x-axis of \figref{fig:figure2}).
{As a result, the control optimality metric proves to be a more sensitive indicator than circuit expressibility to account for highlighting the diversity among the three ansatz. Therefore, the normlized $W_1$ distances for Ansatz-a, Ansatz-b, and Ansatz-c, as calculated from the CDFs in \figref{fig:figure2}(b), are approximately 0.921, 0.689, and 0.2585, respectively. Correspondingly, the estimated KL divergences ($D_{\text{KL}}$) are 12.58, 5.66, and 10.11. In this case, expressibility does not consistant with performance across different control strategies.}

\begin{figure}[h] 
\centering
\includegraphics[width=1\columnwidth ]{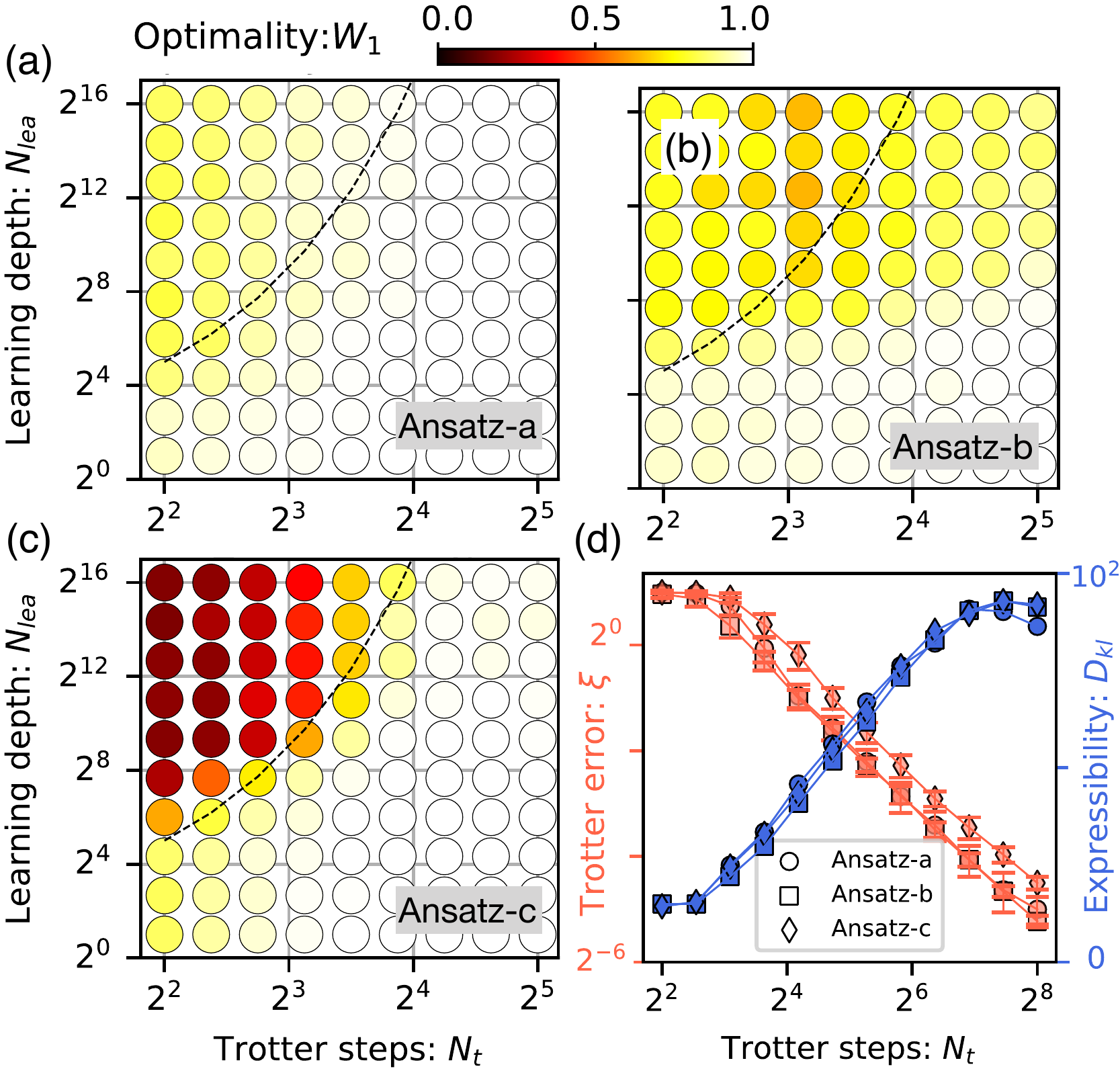}
\caption{
(a-c). The $W_1$ distance as function of $N_{\rm lea}$ and Trotter step $N_t$ for Ansatz-a,-b and -c, respectively.
Note that each data point is calculated over $20$ samples, and the dashed line denotes $N_{\rm lea}\sim \mathcal{O}(2^{N_t})$.
In (d), we present the Trotter error (left y-axis) and $D_{kl}$ (right y-axis) versus the Trotter step for three ansatze illustrated by different shapes (see legend), where the error bar refers the standard deviation over $10^2$ samples.
Parameters: qubit number $N=6$ and total time $T=3$.
}
\label{fig:figure3}
\end{figure}

{Furthermore, we investigate the optimality of the maximum-fidelity control as a function of the Trotter step $N_t$ and learning depth $N_{\rm lea}$. In \figref{fig:figure3}(a-c), we present the $W_1$ distance (cf. Eq.\eqref{eq:W1}) for three ansatz in parametric grid $\{N_{\rm lea},N_t\}$. The corresponding Trotter error $\xi$  (cf. Eq.\eqref{eq:trotter_error}) and KL divergence (cf. Eq.\eqref{KL}) are compared in (d), respectively. 
The results show that circuits with higher correlations exhibit lower performance, consistent with previous studies~\cite{Expressibility2022prxq}, as effectively captured by the $W_1$ distance in Fig.\ref{fig:figure3}(a-c). However, the expressibility metric (represented by the $KL$ divergence and Trotter error in Fig.\ref{fig:figure3}(d)) fails to reflect the practical performance among diverse ansatz. 
{Counterintuitively, increasing the number of Trotter steps, $N_t$, does not necessarily enhance the performance of VQAs. We attribute this phenomenon to the negligible impact of Trotter error compared to its influence on the optimizer, where a larger $N_t$ increase in the number of free variables to be optimized (requiring larger learning budget $\mathcal{O}(2^{N_t})$, see dashed curve in \figref{fig:figure3}), potentially complicating the optimization process. On the other hand, the less Trotter error reduce the expressibility of quantum circuit~\cite{Wurtz2022counterdiabaticity}.}
These findings suggest that control optimality, as quantified by the $W_1$ distance, provides a more consistent and accurate description of VQA performance.
}

Then we can conclude that to evaluate the overall performance of a VQA, one should estimate \textit{control optimality} rather than \textit{expressibility} for the following reasons: \( W_1 \) reflects both the expressibility of the quantum circuit and the performance of the classical optimizer, whereas \( D_{KL} \) only captures the former. This implies that under certain conditions, \( W_1 \) can serve as a substitute for \( D_{KL} \). As well, from a mathematical point of view, when two probability distributions have nearly no overlap, the Wasserstein distance is often a more meaningful and precise measure of their separation compared to \( D_{KL} \).

\subsection{The time-optimal control}
The speed of state propagation in quantum systems is fundamentally bounded by the time-energy relation, which is addressed by the concept of quantum speed limits (QSLs)~\cite{MT45,AA90,Uhlmann92,ML98}. For a time-independent case, there are two expressions for the QSL:
$\tau_{\text{QSL}}\equiv\max\{\frac{\pi\hbar}{2{E}},\frac{\pi\hbar}{2{\Delta E}}\},$
 refer to the Mandelstam-Tamm QSL taking into account energy dispersion $\Delta E$~\cite{MT45} and Margolus-Levitin bound~\cite{ML98} using the mean energy $E$ instead. 
For a generic driven system $\hat{H}(t)$,  only an analog of the Mandelstam-Tamm bound is known ~\cite{AA90,Uhlmann92}. 
From a geometric point of view, the distance between two states illustrated by Bures angle $\mathcal{L}(\psi_0,\psi_t)=\arccos( | \langle \psi_0 | \psi_t \rangle|)\in[0,\pi/2]$.
The maximum speed required to sweep a given Bures angle is quantified by 
QSL reads
$\tau_{\text{QSL}} =\frac{\hbar}{\overline{\Delta E}}\mathcal{L}(\psi_0,\psi_f),$
with the time-averaged energy dispersion
$
\overline{\Delta E} = \frac{1}{T}\int_0^Tdt\sqrt{\langle \psi_t |\hat{H}(t)^2|\psi_t\rangle-\langle \psi_t |\hat{H}(t)|\psi_t\rangle^2}.
$
In this manner, the QSL is approached by maximizing the energy dispersion at all time, while keeping the system evolved to the target states.  
For in-depth studies
of these, we refer the readers to  and the reference therein~\cite{deffner2013energy,Deffner17rev}.

Notably, the time-optimal control is the minimum time required to reach the unit fidelity for a quantum control  task, which is bounded by the QSL, i.e. $\tau_{\text{QSL}}\le T_{\text{min}}$, since the controller nevertheless constrained by physical condition or optimization hardness.
The QSLs of PST in a $N$-spin chain is usually interpreted as the  \textit{time-per-site} over $N-1$ orthogonal swaps between two neighboring sites since there is no direct interaction between initial and final site.
Therefore, the QSL can be quantified by $\tau_{\text{QSL}} = (N-1)\tau^*_{\text{QSL}}$, where $\tau^*_{\text{QSL}}=\pi/(4J_0)$ is the QSL from one site to neighboring site.
In previous studies~\cite{MH06pra,kay2022notespeedperfectstate,Nori12pra}, the minimum time to PST is approximately
\beq\label{time_minimum}
T_{\text{min}} = \frac{N-1}{2J_0},
\eeq
where $J_0$ is coupling strength. More generally, for different control strategies, one can quantify the speed limit by a linear function of spin number~\cite{krotov2011pra,ocqsl09prl}
\beqa\label{h}
T^*_{\text{min}} = h(N-1)/J_0,
\eeqa
where the $h$ is a dimensionless constant.
 To reach this limit, instead of the conventional algorithm~\cite{krotov2011pra,ocqsl09prl,RLzhang2018}, we utilize the VQA-based method relied on three previously defined ansatz.

\begin{figure}[h]
\centering
\includegraphics[width=0.9\columnwidth ]{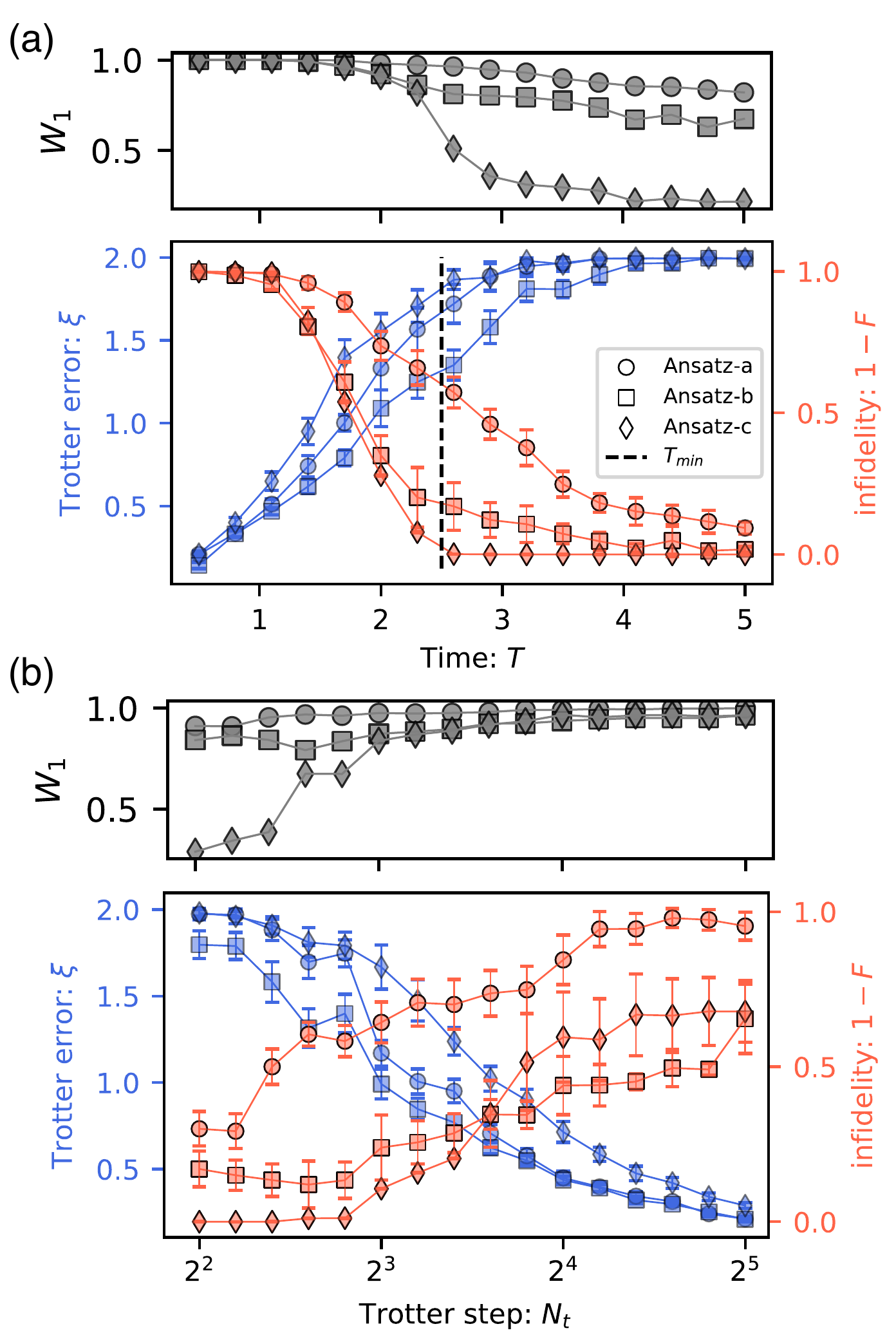}
\caption{ The Trotter error (left y-axis) and infidelity (right y-axis) versus the time and Trotter step in (a-b), respectively. The vertical dashed line refers to the $T_{\rm min}=(N-1)/2$. And the $W_1$ distance is accordingly presented on the upper panel in (a-b), respectively.
Note $N_t=N$ in (a), and $T=T_{\rm min}$ in (b), and the error bar is the standard deviation of $10^2$ samples.
Other parameters are the same as those in \figref{fig:figure3}.
}
\label{fig:figure4}
\end{figure}

{In \figref{fig:figure4}, we present the performance of VQA-based time-optimal control for three ansatz, respectively. The optimality $W_1$, Trotter error $\xi$ and infidelity ($1-F$) are illustructed for the maximum-fidelity control with varying total time $T$ in (a) and Trotter step $N_t$ in (b). Note that we chose $N_t=6$ in (a), and $T=T_{\rm min}$ (dashed line in (a)) in (b).
As a result, the minimum-time control of PST is approximatly $2.52$ which is closed to the anlytic estimation (see Eq.\eqref{time_minimum}) while the fidelity fullfills $F>{F}_c = 0.999$ by using the uncorrelated-ansatz-based (Ansatz-c) VQA method. For Ansatz-a, the minimum-time for reaching fidelity $F\ge F_c$ is approximately $T\ge 4$ which is larger than the case with Ansatz-c. The fidelity of the most correlated Ansatz-a is converging slowly than other two ansatz as shown in the \figref{fig:figure4}(a). Similar result is shown in \figref{fig:figure4}(b) when we set Trotter steps as the variable (given $T=T_{\rm min}$), which means that small Trotter error does not meets the high performance in this case. We interpret this counter-intuitive phenomone is that the Trotter error can be regarded as an extension of optimization landscape, resulting a higher expresibility (see \figref{fig:figure3}(c)).
Note similar observation also discussed in Ref.~\cite{Wurtz2022counterdiabaticity}. 
It is shown that the VQA-based maximum-fidelity control is highly constrained by the given time and ansatz correlation, which can be reflected by the behavior of $W_1$ distance (see upper panel of \figref{fig:figure4}(a-b)). Hereafter, we take Ansatz-c for the furture analysis as it yields the favorable performance among three ansatz based on the results shown in \figref{fig:figure3} and \figref{fig:figure4}.}

\begin{figure}[h]
\centering
\includegraphics[width=0.85\columnwidth ]{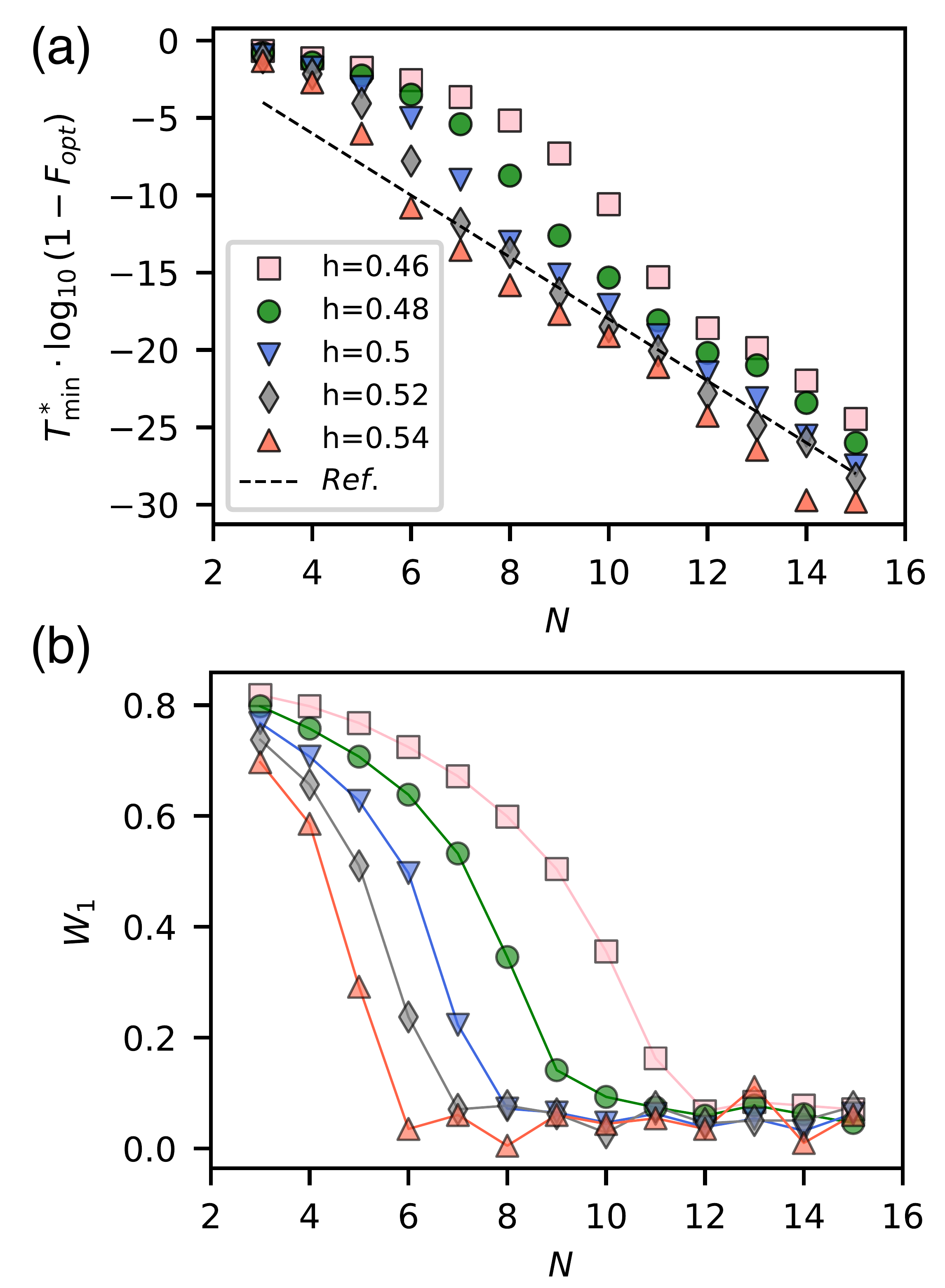}
\caption{(a). The optimal control ($F_{\rm opt}\cdot T^* $) as a function of qubit number $N$, where the Trotter step $N_t=N$ and $T^*=h(N-1)$ for different h in legend. Each data point is averaged our by $10^2$ numerical optimization, and the dashed line refers $h=0.5$ and $F_{\rm opt}=1-10^{-4}$. (b). The  $W_1$ distance of corresponding $10^2$ numerical results in (a) with respect to $N$.  
}\label{fig:figure5}
\end{figure}

{In \figref{fig:figure5}(a), we present the time-optimal control ($\log_{10}(1-F_{\rm opt})\cdot T^* $) by using Ansatz-c-based VQA as a function of qubit number $N$ for different scaling factor, i.e., $h=\{0.46,0.48,0.5,0.52,0.54\}$, which determines the total time. Note that the optimal fidelity $F_{\rm opt}$ refers the fidelity averaged out $10^2$ numerical experiments, and the dashed line represents when $h=0.5$ and fidelity $F_{\rm opt}=1-10^{-4}$. The distance between each datapoint from the reference line denotes the performance of time-optimal control including the fidelity and total time, which can be reflected by the corresponding $W_1$ distance in (b). 
One can see that our method acquires tight time limit $(h\le 0.5)$ at larger spin number $N>9$ while the  infidelity remains $ \sim 10^{-4}$ ($W_1\to 0$). This result only requires Trotter step as a linear function of qubit number, i.e., $N_t=N$ in \figref{fig:figure5}, indicating the practical implementation in near-term quantum computer by considering the circuit depth.}

\begin{figure} 
\centering
\includegraphics[width=0.85\columnwidth]{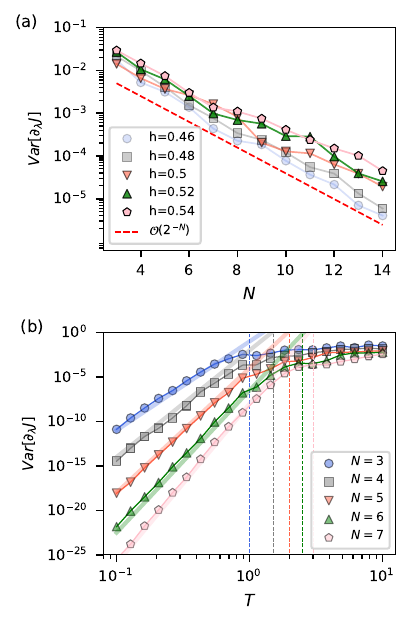}
\caption{
 (a) The variance of the control gradient, defined in Eq.~\eqref{grad_est}, is shown as a function of the qubit number $N$, with the total time $T = h(N-1)/J_0$, where $h \in [0.46, 0.48, 0.5, 0.52, 0.54]$ and $N_t = N$. The red dashed line represents the scaling $\mathcal{O}(2^{-N})$.
 (b) The gradient variance as a function of total time $T$ is plotted for qubit numbers $N \in [3, 4, 5, 6, 7]$ (marked solid lines). The corresponding slope lines indicate the scaling $\mathcal{O}(T^{2N})$, while the vertical dashed lines mark the minimum time $T_{\text{min}} = (N-1)/(2J_0)$ with $N_t = 2N$.
}
\label{fig:figure6}
\end{figure}

The gradient behavior, including its scaling with the number of qubits $N$ and circuit depth $N_t$, is crucial for evaluating the performance of gradient-based hybrid quantum algorithms. \ For instance, estimating the fidelity of $N$-qubit states through standard quantum state tomography requires measurements that scale as $\mathcal{O}(4^{2N})$~\cite{QST}, making it prohibitively expensive. This exponential cost significantly limits the practical applicability of gradient-based optimizers in variational quantum algorithms. 
In our study, we analyze the control gradient derived from Eq.\eqref{grad_est} in our setup. 
In \figref{fig:figure6}, we illustrate the scaling behavior of the gradient magnitude and variance under different conditions. \figref{fig:figure6}(a) presents the variance of the gradient as a function of the qubit number $N$, with circuit depth (Trotter step) scaled as $\mathcal{O}(N)$. For different values of the parameter $h$ (total time in Eq.\eqref{h}), the gradient variance exhibits an approximate scaling of $\mathcal{O}(2^{-N})$ (shown by the dashed red line), which is significantly less steep than the previously reported $\mathcal{O}(2^{-2N})$ scaling~\cite{mcclean2018barren}. This phenomenon is due to the highly correlated ansatz compared to the correlation-free random circuit~\cite{mcclean2018barren,Expressibility2022prxq}.  

In \figref{fig:figure6}(b), we show the gradient variance as a function of the total time $T$ for various qubit numbers $N$.
In the short-time regime $(T \ll T_{\text{min}})$, the gradient variance decreases exponentially, following an empirical scaling of 
\beq\label{eq:var_T}
\text{Var}[\partial_\lambda J]\sim\mathcal{O}(T^{2N}) \quad \text{while} \quad (T \ll T_{\text{min}}),
\eeq
where the circuit depth is a linear function of qubit number, e.g., $N_t = N$ in \figref{fig:figure6}.
As the evolution time approaches the minimum value $T_{\text{min}} = (N-1)/2$, the scaling transitions to a constant. In comparison, we refer vertical dashed lines to the respective $T_{\text{min}}$ values for different qubit numbers. 
This behavior highlights a key feature of the speed limit in time-optimal control problems, where the gradient variance is vanishing at shorter timescales ($T\le T_{\text{min}}$). At these timescales, achieving high-fidelity solutions becomes difficult, leading to poor control optimality ($W_1 \to 1$), which is consistent with the results shown in Fig.\ref{fig:figure4}.

\section{Conclusion and Outlooks}
  
 \textit{Conclusion}. {In this study, the quantum-classical hybrid method provides an alternative path for addressing optimal control problems in many-body quantum systems. In specific, we demonstrate the application of VQAs to achieve time-optimal control for perfect-state transfer in a spin chain system. The results show that the VQA with an uncorrelated ansatz approaches the speed limit, performing even better than previous conventional methods ($h<0.5$ in \figref{fig:figure5}).  To systematically evaluate the overall performance of hybrid workflows, we introduce a metric called \textit{control optimality} quantified by the $W_1$ distance of the resultant fidelity density distrubution from a target one, which integrates effects from both quantum circuits and classical optimizer performance. 
 We numerically benchmark the behavior of $W_1$ distance with respect to the circuit ansatz (circuit expresibility), circuit depth (Trotter errors), and total time (physical constrain).
 }

We also analyze the trainability of VQA from the perspective of classical optimization. Specifically, the gradient variance was examined as a function of the number of qubits for linearly increasing the depth of the circuit. Contrary to previous observations where the gradient variance scales as $\mathcal{O}(2^{-2N})$~\cite{mcclean2018barren}, our results reveal a scaling of $\mathcal{O}(2^{-N})$, attributed to deterministic layers that prevent the circuit from forming a 2-design~\cite{Expressibility2022prxq}. Furthermore, we observed a significant reduction in the gradient, scaled as $\mathcal{O}(T^{2N})$, when the evolution time deviated from the speed limit ($T\ll T_{\rm min}$). This highlights the physical limitations on trainability, suggesting that gradient vanishing occurs when controllability is insufficient for time-optimal tasks within a fixed evolution time below the quantum speed limit. Moreover, we study the error tolerance of our method  in the presence of state-preparation-and-measurement error, see the Appendix \ref{app3}. 
In addition, uncertainties in cost values caused by SPAM errors can be mitigated using techniques such as randomized benchmarking~\cite{RB} and machine learning-assisted protocols~\cite{palmieri2020experimental,Tangyou22PrA}.

In summary, we demonstrate a hybrid quantum-classical approach for solving control problems in many-body quantum systems. In our case study, we achieved time-optimal control for perfect state transfer in a spin chain, using a shallow circuit scaled linearly with qubit number. We systematically addressed challenges from both quantum and classical perspectives with a new metric termed control optimality, emphasizing the robustness of our method to Trotter errors, making it suitable for near-term quantum processors.

\textit{Outlook.}
{Firstly, from the perspective of experimental realization, this method can be extended to a pulse-level optimization~\cite{magann2021pulses,PhysRevApplied.22.024009,li2022pulse}, where gate operations are replaced by optimized pulse parameters, such as envelope shapes and frequencies, providing a more experimentally feasible approach. 
Secondly, our proposed \textit{control optimality}, as a general metric, statistically reflects the overall performance of a hybrid quantum-classical framework, which can be directly extended to performance evaluation and predictive analysis for a VQA~\cite{cerezo2021variational}, quantum machine learning~\cite{cerezo2022challenges} and closed-loop quantum optimal control~\cite{QOC2014RB}.
We leave this extension to future works.
Third, these findings offer valuable insights into leveraging hybrid quantum-classical algorithms for quantum control and demonstrate the potential for achieving practical advantages over classical counterparts. 
For example, this method can be implemented to optimally prepare the Mott insulator state in a Bose-Hubbard model~\cite{PhysRevLett.106.190501}, fast magnon transport in large spin limits compared with the invariant-based Shortcuts to Adiabaticity~\cite{ magnon2021fast,magnon2022optimal} and other quantum control of many-body system~\cite{PhysRevLett.121.040503,PhysRevA.88.012334}.
}

\section{Acknowledgments}
T.H thank to L.D and W.R for useful discussion.
This work is fully supported by CPS-MindSpore funding. T. H. thanks to the technical team of MindSpore Quantum for their support.
The numerical calculations in this work were mainly performed using the MindSpore Quantum library \cite{mq_2021}. The data and codes that support the findings of this study are openly available at {repository} (~\href{https://gitee.com/mindspore/mindquantum/tree/research/paper_with_code}{{https://gitee.com/mindspore/mindquantum/tree/research/\\paper\_with\_code}}).

\appendix

\section{Circuit construction}\label{app1}
Following the procedure outlined for VCL in the previous section, we first prepare the initial states by applying an ${\tt X}$ gate on the first qubit. This operation can be experimentally realized using an enhanced $\pi$-pulse (e.g., a DRAG pulse) with a fidelity exceeding $99.99\%$~\cite{PRXQuantum.5.040342}. Next, we construct the time-dependent propagator of the Hamiltonian in Eq.\eqref{Heisenberg} using Trotter decomposition. Specifically, the propagator $\textbf{U}{\text{XXZ}}(0,T)$ is expressed as $\prod{j=1}^{N_t} {U}_{\text{XXZ}}(t_j)$, where the time-varying parameters are discretized into a sequence ${0, \Delta t, 2\Delta t, \dots, N_t\Delta t}$ with $\Delta t = T/(N_t-1)$.
For each Trotter step, we have
\beq
{U}_{\text{XXZ}}(t_j) = R^{j}_{XY} R_Z^{j},
\eeq
where $R_{XY}[\alpha_{n,n+1}(t_j)]$ is a neighboring SWAP-type two-qubit gate and $R_Z^{j}$ is a single-qubit rotation gate. The two-qubit gate is expressed as: $R_{XY}[\alpha_{n,n+1}(t_j)] =e^{-i\alpha_{n,n+1}(\sigma^x_{n}\sigma^x_{n+1}+\sigma^y_{n}\sigma^y_{n+1})}$, where $\alpha_{n,n+1} = -J_0 \Delta t/2$ is a deterministic coefficient. The single-qubit gate is given by
$R_Z^j[\theta_n(t_j)] = e^{-i\theta_n(t_j)\sigma^z_n}$,
where $\theta_n(t_j)$ contains trainable parameters, and $n$ denotes the qubit index (see Eq.\eqref{Heisenberg} and Eq.\eqref{f_fun}). In this work, we construct the gate-based circuit using a noise-free simulator, deferring the discussion of noisy environments to a later section.

For initialization, we randomly set the controller $d(0)\in[-N,N]$ $(A_0=J_0)$ for the \textit{Ansatz-a} ansatz, and $d(0)\in[-N,N]$, $A(0)\in[-J_0,J_0]$ for the \textit{Ansatz-b} ansatz. For the \textit{Ansatz-c} ansatz, the initial parameters $\theta_n(0)$ are sampled randomly from $\theta_n(0)\in[0,2\pi J_0]$. Therefore, the trainable gates ${R_Z^{j}(\theta_n)}_j$ are updated iteratively using a classical optimizer in a closed-loop optimization process.
The number of variables to be optimized in the trainable layer scales approximately as $\sim\mathcal{O}(N_t)$, $\sim\mathcal{O}(2N_t)$, and $\sim\mathcal{O}(N\cdot N_t)$ for the \textit{Ansatz-a}, \textit{Ansatz-b}, and \textit{Ansatz-c} ansatze, respectively. For simplicity, all variables (time and energy) are scaled with respect to ${J_0}$.
Moreover, we randomly initialize the controller $d(0)\in[-N,N]$ $(A_0=J_0)$ for the \textit{Ansatz-a} ansatz, and $d(0)\in[-N,N]$,$A(0)\in[-J_0,J_0]$ for \textit{Ansatz-b} ansatz. We randomly sample $\theta_n(0) \in[0,2\pi J_0]$ for the initial set of \textit{Ansatz-c} ansatz.
Therefore, the trainable gate $\{R_Z^{j}(\theta_{n})\}_j$ can be updated according to the classical optimizer in a closed-loop optimization manner. 
In explicit, the number of variables to be optimized in the trainable layer is scaled as $\mathcal{O}(N_t)$, $\mathcal{O}(2 N_t)$ and $\mathcal{O}(N\cdot N_t)$ for above three ansatz, respectively.

\section{The bound of optimality}\label{app2}
In this appendix, we bound the 1-Wasserstein distance for quantifying the optimality of the variational circuit learning.
In our setup, we consider  the 1-Wasserstein ($W_1$) distance $W_1[P, Q]$ between two probability distributions, e.g., $P(\mathcal{X})$ and $Q(\mathcal{X})$, on the a given basis  $\mathcal{X} = \{x_0,dx,2dx,...,(\mathcal{N}-1)dx,x_f\}$ divide by $dx =\Delta x/(\mathcal{N}-1)$ with $\Delta x= x_f-x_i$. 
According to the definition of 1-Wasserstein distance, we have the lower bound $ W_1[P, Q] =0$ while distribution $P$ and $Q$ are identical. 
{In other hands, we define the distance function $d(x_i,x_j) =|i-j|\times dx$ satisfying the symmetry condition $W_1(P,Q)\equiv W_1(Q,P)$, and the maximum value $W_1[P, Q] =  \Delta x$ while the $P(\mathcal{X}) = \delta_{i0} $ and $Q(\mathcal{X}) = \delta_{i \mathcal{N}} $ are delta function at edges $P(x_0)=1$ and $Q(x_N)=1$.} Therefore, for a given basis $\mathcal{X} $, the $W_1$ distance of two distribution is bounded by
\beq
0\le W_1(P(x),Q(x))\le  \Delta x.
\eeq

\begin{figure}\label{fig:figure7}
\centering
\includegraphics[width=1\columnwidth ]{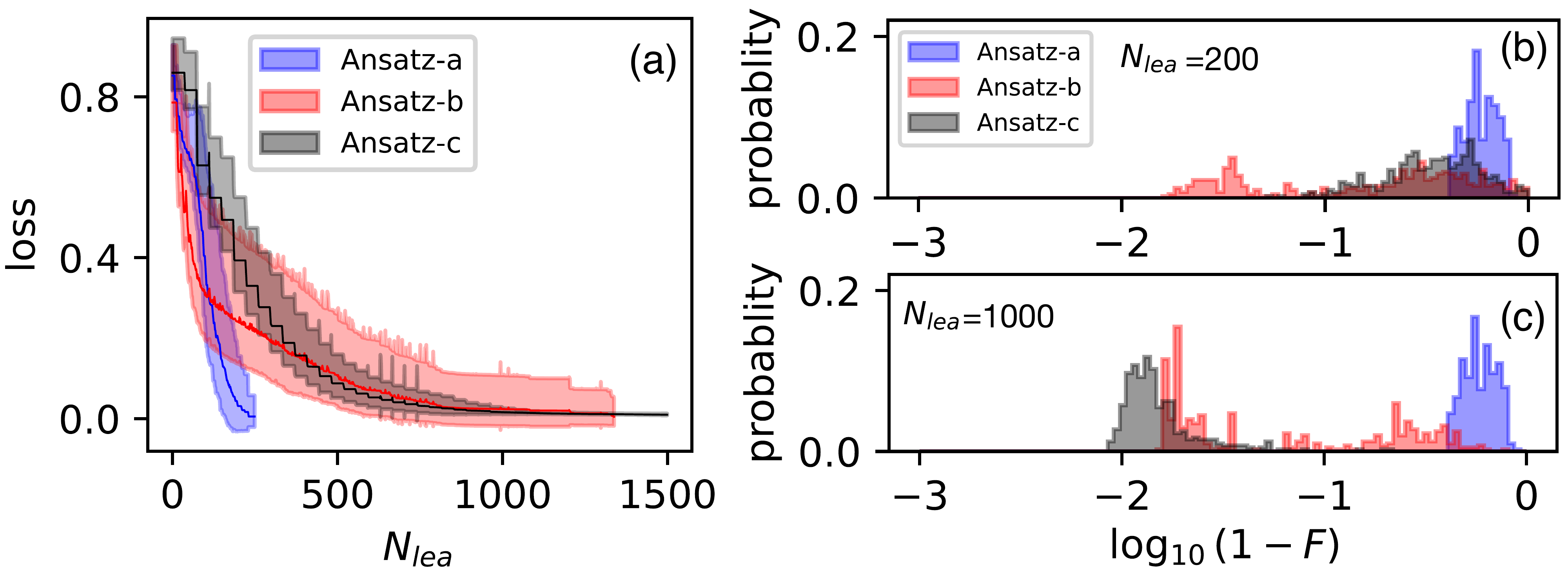}
\caption{{(a) The loss value, defined as infidelity $1-F$, is plotted as a function of the number of loss evaluations $N_{\mathrm{lea}}$ for three different ansätze, with all other settings kept fixed. The solid lines represent the average over $10^2$ numerical experiemnts, with the standard deviation indicated by the shaded area. Two representative cases from (a), corresponding to $N_{\mathrm{lea}} = 200$ and $N_{\mathrm{lea}} = 1000$, are shown in panels (b) and (c), respectively. Other parameters are the same as in \figref{fig:figure2}.}}
\end{figure}

{Next, we discuss the convergence of the $W_1$ distance. Here, convergence refers to the case where $N_{\rm lea} \to \infty$, meaning each optimization process converges to a local minimum. First, note that the learning cost in our setting can take a finite value, and different convergence outcomes may lead to variations in the evaluation of VQA performance. For example, in \figref{fig:figure7}(b-c), it is evident that when $N_{\rm lea} = 1000$, the computed $W_1$ values exhibit greater discrepancy among diverse ansatze compared to the case when $N_{\rm lea} = 200$. In other words, when using the $W_1$ distance to assess VQA performance, the value of $N_{\rm lea}$ (i.e., the degree of convergence) affects the sensitivity of the $W_1$ measure.}


\section{Error robustness}\label{app3}
\begin{figure} 
\centering
\includegraphics[width=0.92\columnwidth]{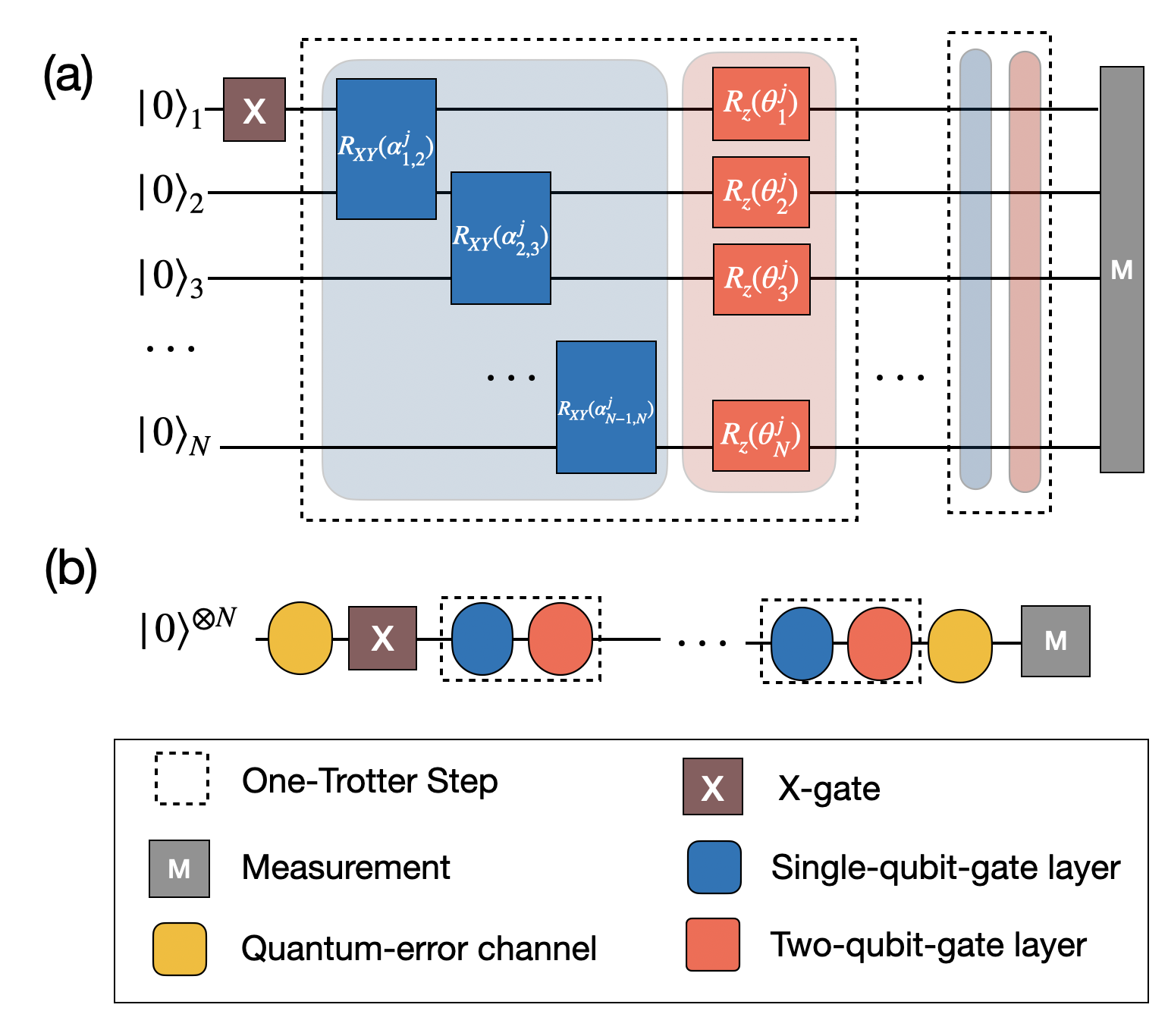}
\caption{\textbf{The layout of gate-based quantum circuit.}
(a). The $N$-qubit quantum circuit for Trotter decomposition of Hamitonian \eqref{Heisenberg}, including $X$ gate, $XY$-SWAP-typed two-qubit gate, and $Z$-rotating gate. In (b), we present the SPAM error simulation by introducing the error channel (pink block) after qubit initialization and before measurement.  Moreover, we mimic the gate error by adding error channel after each Trotter layer.
}
\label{figure7}
\end{figure}

In state-of-the-art experiments, one- and two-qubit gate fidelity have reached error levels below 0.1\%~\cite{PhysRevLett.123.210501,PhysRevLett.125.120504}. 
Notably, the two-qubit ${R_{XY}}$ gate, incorporating the $XY$ exchange interaction, can be directly implemented by dispersively coupling two transmon qubits to the same resonator~\cite{Heras2014prl}. Experimentally, such SWAP-type two-qubit gates have been realized in superconducting circuits with fidelities exceeding 99.9\% ~\cite{PhysRevLett.123.210501}, so-called \text{fSim} gate ~\cite{PhysRevLett.125.120504}. The circuit layout and gate representation are illustrated schematically in Fig.~\ref{figure7}.
On the other hand, hardware noise, such as decoherence and crosstalk, is unavoidable in near-term quantum devices. 
However, errors in qubit initialization and measurement, commonly referred to as state preparation and measurement (SPAM) errors, are typically at the 1\% level~\cite{PhysRevLett.133.170601}. 
In the context of VQAs, noise may slow down the training process, bias the optimization landscape, and shift the global optimum of the cost function~\cite{cerezo2021variational}.
From an experimental perspective, SPAM errors significantly hinder the performance of VQA-based quantum optimal control. 
 To address the error robustness of our framework, we here only consider the uncertainty induced by SPAM errors, modeled using well-defined quantum channels~\cite{nielsen2002quantum}. 

\begin{figure}[h]
\centering
\includegraphics[width=0.7 \columnwidth]{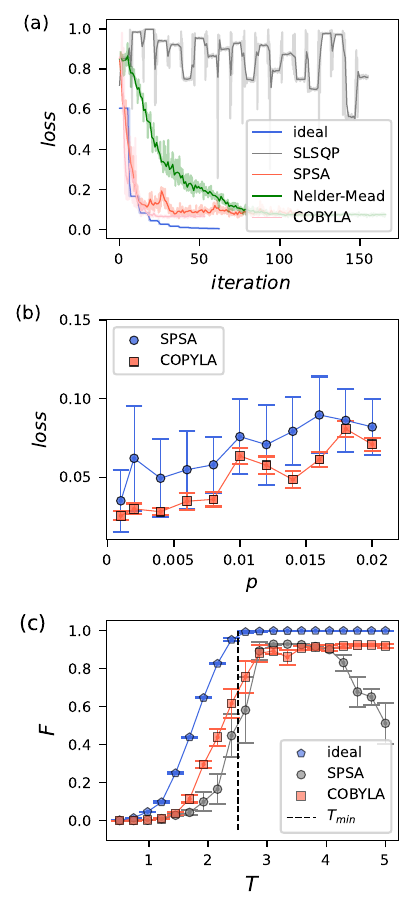}
\caption{\textbf{The VQA performance in a noisy environment}.
(a) Loss function convergence over iterations for various optimizers, where ${\tt COPYLA}$ demonstrates stable and rapid convergence compared to other methods.
(b) Average loss over the last ten iterations of the VCL optimization process as a function of depolarizing error strength $p$, showing that ${\tt COPYLA}$ consistently outperforms ${\tt SPSA}$. The control tasks in (a) and (b) are identical to those in the \figref{fig:figure3}. (c) Fidelity $F$ versus total time $T$ at $p=0.01$, comparing results obtained from ideal (noiseless) and noisy circuits.
Error bars in (b) and (c) represent one-sigma uncertainty based on ten numerical experiments. Other setups in (c) are identical to those in the \figref{fig:figure4}.
}
\label{figure8}
\end{figure}

For numerical experiments, we incorporate depolarizing noise modeled as: 
\beq E(\rho) = (1-p)\rho + \frac{p}{2^N} \text{Tr}[\rho]I, 
\eeq 
where $p \in [0, 4^N/(4^N-1)]$. Here, $p = 0$ corresponds to the identity channel, and $p = 4^N/(4^N-1)$ represents the fully depolarized channel. To mimic SPAM errors, we introduce an error layer associated with qubit initialization and measurement, as shown in Fig.~\ref{figure8}. As an example, we perform the VQA-based control task under the same conditions as in Fig.\ref{figure8}, but now on a noisy circuit.  

In Fig.\ref{figure8}(a), we present the convergence of the loss function over iterations on a noisy simulator ($p=0.01$) for various optimization strategies, including gradient-based methods (${\tt SLSQP}$, ${\tt SPSA}$), gradient-free methods (${\tt Nelder-Mead}$, ${\tt COBYLA}$), and the ideal case. The results demonstrate that SPAM errors can significantly limit the training process and may lead to convergence at local minima, rendering regular gradient-based optimizers like ${\tt SLSQP}$~\cite{SLSQP} ineffective. In contrast, gradient-free optimizers, specifically ${\tt COBYLA}$~\cite{powell1998direct} and ${\tt SPSA}$~\cite{spall1998overview}, exhibit rapid and stable convergence under noisy conditions. Other optimizers either exhibit fluctuations or slower convergence rates.

In Fig.~\ref{figure8}(b), we benchmark the performance of ${\tt SPSA}$ and ${\tt COBYLA}$ by plotting the loss value (infidelity) as a function of error strength $p \in [0, 0.02]$. Notably, the infidelity converges to approximately $10^{-2}$ for the noise-free case ($p = 0$), which aligns with the measurement precision $\sim 1/\sqrt{N_r}$, where the shot number $N_r = 4096$ in our simulation. As the error strength increases, ${\tt COBYLA}$ maintains lower loss values compared to ${\tt SPSA}$, highlighting its better robustness in noisy environments.

Finally, in Fig.\ref{figure8}(c), we evaluate the fidelity $F$ achieved by VCL as a function of the total evolution time $T$. The results reveal that ${\tt COBYLA}$ demonstrates greater robustness to SPAM errors compared to ${\tt SPSA}$, especially near the theoretical minimum time $T_{\text{min}}$ given in Eq.\eqref{time_minimum}. However, in the presence of SPAM errors, the optimal time is shifted, indicating the noise-induced modification of the theoretical speed limit.

\newpage
\bibliography{ref}	
\bibliographystyle{apsrev4-2}
\end{document}